\documentclass[11pt,english]{article}
\usepackage[T1]{fontenc}
\usepackage[latin9]{inputenc}
\usepackage[a4paper]{geometry}
\usepackage{amsmath}
\usepackage{setspace}
\usepackage{amssymb}
\makeatletter
\newcommand{\lyxaddress}[1]{
\par {\raggedright #1
\vspace{1.4em}
\noindent\par}
}

\makeatother

\usepackage{babel}

\begin{document}

\title{\textbf{Generalization of Schwinger-Zwanziger Dyon to Quaternion}}

\author{O. P. S. Negi$^{\text{(1, 2)}}$%
\thanks{\textbf{Address from November 08- December 22, 2010:- Universität
Konstanz, Fachbereich Physik, Postfach-M 677, D-78457 Konstanz, Germany}%
}, H. Dehnen$^{\text{(1)}},$ Gaurav Karnatak$^{\text{(2)}}$ and P.
S. Bisht$^{\text{(2)}}$}

\maketitle

\lyxaddress{\begin{center}
$^{\text{(1)}}$Universität Konstanz\\
 Fachbereich Physik\\
 Postfach-M 677\\
 D-78457 Konstanz, Germany
\par\end{center}}

\lyxaddress{\begin{center}
$^{\text{(2)}}$Department of Physics,\\
Kumaun University,\\
S. S. J. Campus, \\
Almora-263601(Uttarakhand) India
\par\end{center}}

\lyxaddress{\begin{center}
Email- ops\_negi@yahoo.com\\
Heinz.Dehnen@uni-konstanz.de\\
gauravkarnatak2009@yahoo.in\\
ps\_bisht123@rediffmail.com
\par\end{center}}
\begin{abstract}
Postulating the existence of magnetic monopole in electromagnetism
and Heavisidian monopoles in gravitational interactions, a unified
theory of gravi-electromagnetism has been developed on generalizing
the Schwinger-Zwanziger formulation of dyon to quaternion in simple
and consistent manner. Starting with the four Lorentz like forces
on different charges, we have generalized the Schwinger-Zwanziger
quantization parameters in order to obtain the angular momentum for
unified fields of dyons and gravito-dyons (i.e. Gravi-electromagnetism).
Taking the unified charge as quaternion, we have reformulated manifestly
covariant and consistent theory for the dynamics of four charges namely
electric, magnetic, gravitational and Heavisidian associated with
gravi electromagnetism. 

\textbf{PACS No: 14.80 Hv.}
\end{abstract}

\section{Introduction}

~~~~ The question of existence of monopole \cite{key-1} and dyons
\cite{key-2} has become a challenging new frontier and the object
of more interest in high energy physics. Dirac showed \cite{key-1}
that the quantum mechanics of an electrically charged particle of
charge $e$ and a magnetically charged particle of charge $g$ is
consistent only if $eg=2\pi\, n$, $n$ being an integer. Schwinger-Zwanziger
\cite{key-2} generalized this condition to allow for the possibility
of particles (dyons) that carry both electric and magnetic charge.
A quantum mechanical theory can have two particles of electric and
magnetic charges $(e_{1},\, g_{1})$ and $(e_{2},\, g_{2})$ only
if $e_{1}g_{2}-e_{2}g_{1}=2\pi\, n$. The angular momentum in the
field of the two particle system can be calculated readily with the
magnitude $\frac{e_{1}g_{2}-e_{2}g_{1}}{4\pi c}$. This has an integer
or half-integer value, as expected in quantum mechanics, only if $e_{1}g_{2}-e_{2}g_{1}=2\pi n\hslash c$.
The fresh interests in this subject have been enhanced by 't Hooft
-Polyakov \cite{key-3} with the idea that the classical solutions
having the properties of magnetic monopoles may be found in Yang -
Mills gauge theories. Julia and Zee \cite{key-4} extended the ' t
Hooft-Polyakov theory \cite{key-3} of monopoles and constructed the
theory of non Abelian dyons. The quantum mechanical excitation of
fundamental monopoles include dyons which are automatically arisen
from the semi-classical quantization of global charge rotation degree
of freedom of monopoles. In view of the explanation of CP-violation
in terms of non-zero vacuum angle of world \cite{key-5}, the monopoles
are necessary dyons and Dirac quantization condition permits dyons
to have analogous electric charge. Accordingly, a self consistent
and manifestly covariant theory has been developed \cite{key-6} for
the generalized electromagnetic fields of dyons. 

On the other hand, the analogy between linear gravitational and electromagnetic
fields leads to the asymmetry in Einstein\textquoteright{}s linear
equation of gravity and suggests the existence of gravitational analogue
of magnetic monopole \cite{key-7}. Cattani \cite{key-7} has also
derived the covariant field equations for linear gravitation like
Maxwell's equations on introducing a new (called the Heavisidian )
field ( like magnetic field ) which depends upon the velocity of gravitational
charges (masses). Avoiding the use of arbitrary string variable, the
manifestly covariant and consistent theory of gravito - dyons has
also been developed \cite{key-8} in terms of two four-potentials
leading to the structural symmetry between generalized electromagnetic
fields of dyons and generalized gravito-Heavisidian fields of gravito-dyons.
Extending this recently, a consistent theory for the dynamics of four
charges (masses) (namely electric, magnetic, gravitational, Heavisidian)
have also been formulated \cite{key-9} in simple, compact and consistent
manner. Considering an invariant Lagrangian density and its quaternionic
representation, the consistent field equations for the dynamics of
four charges have already been derived \cite{key-10} and it has been
shown that the present reformulation reproduces the dynamics of individual
charges (masses) in the absence of other charge (masses) as well as
the generalized theory of dyons (gravito - dyons) in the absence gravito
- dyons (dyons).

Postulating the existence of magnetic monopole in electromagnetism
and Heavisidian monopoles in gravitational interactions, in the present
paper, a unified theory of gravi-electromagnetism has been developed
on generalizing the Schwinger-Zwanziger formulation of dyon to quaternion
in simple and consistent manner. Starting with the four Lorentz like
forces on different charges, we have generalized the Schwinger-Zwanziger
quantization parameters in order to obtain the angular momentum for
unified fields of dyons and gravito-dyons (i.e. Gravi-electromagnetism).
Taking the unified charge as quaternion, we have reformulated manifestly
covariant and consistent theory for the dynamics of four charges namely
electric, magnetic, gravitational and Heavisidian associated with
gravi electromagnetism. It has been shown that the combined theory
describes the interaction of particles in terms of four coupling parameters
which we name as the different chirality parameters associated to
electric, magnetic, gravitational, and Heavisidian charges. On applying
the various quaternion conjugations, it has also been emphasized that
the combined theory of gravitation and electromagnetism reproduces
the dynamics of generalized electromagnetic (gravito-Heavisidian)
fields of dyons (gravito-dyons) in the absence generalized gravito
Heavisidian (electromagnetic) fields of gravito-dyons( dyons) or vice
versa.

\section{Schwinger-Zwanziger Dyon in Electromagnetic Fields}

	Let us consider the case of a Schwinger-Zwanziger dyon which is described
as a particle of mass $M$ carrying simultaneously the electric charge
$e_{1}$ and magnetic charge. Let this particle moves with velocity
$\overrightarrow{v}$ so that the force $\overrightarrow{F_{e}}$
experienced by an electric charge is described as, 

\begin{eqnarray}
\overrightarrow{F_{e}} & =M\,\frac{d\overrightarrow{v}}{dt} & =e_{1}\left[\overrightarrow{E}+\overrightarrow{v}\times\overrightarrow{H}\right]\label{eq:1}\end{eqnarray}
where $\overrightarrow{E}$ and $\overrightarrow{H}$ are respectively
the the electric and magnetic fields and let us take throughout out
the notation the system of natural units for which $c=\hslash=1$.
Postulation of existence of magnetic monopoles immediately leads to
the the following set of symmetric generalized Dirac Maxwell's (GDM)
equations\cite{key-1,key-2,key-6,key-8,key-9,key-10},

\begin{align}
\overrightarrow{\nabla}\cdot\overrightarrow{E} & =\rho_{e}\nonumber \\
\overrightarrow{\nabla}\cdot\overrightarrow{H}= & \rho_{m}\nonumber \\
\overrightarrow{\nabla}\times\overrightarrow{E} & =-\frac{\partial\overrightarrow{H}}{\partial t}-\overrightarrow{j_{m}}\nonumber \\
\overrightarrow{\nabla}\times\overrightarrow{H}= & \overrightarrow{j_{e}}+\frac{\partial\overrightarrow{E}}{\partial t}\label{eq:2}\end{align}
where $\rho_{e}$ and $\rho_{m}$are respectively the electric and
magnetic charge densities while $\overrightarrow{j_{e}}$ and $\overrightarrow{j_{m}}$
are respectively the electric and magnetic current densities. These
GDM equations are invariant not only under Lorentz and conformal transformations
but also invariant under the following duality transformations between
electric $\mathcal{E}$ and magnetic $\mathcal{H}$ quantities i.e.

\begin{align}
\mathcal{E}\Longrightarrow & \mathcal{E}\cos\vartheta+\mathcal{\mathcal{H}\sin\vartheta}\nonumber \\
\mathcal{H}\Longrightarrow & \mathcal{H}\cos\vartheta-\mathcal{\mathcal{E}\sin\vartheta}\label{eq:3}\end{align}
where $\mathcal{E}=\left(e_{1},\overrightarrow{E},\rho_{e},\overrightarrow{j_{e}}\right)$
and $\mathcal{\mathcal{H}=}\left(g_{1},\overrightarrow{H},\rho_{m},\overrightarrow{j_{m}}\right)$.
For a particular value of $\vartheta=\frac{\pi}{2}$, equations (\ref{eq:3})
reduces to

\begin{align}
\mathcal{E}\longmapsto & \mathcal{H}\,\,\,\,\,\,\mathcal{H}\longmapsto-\mathcal{E}.\label{eq:4}\end{align}
Accordingly the force experienced by a magnetic monopole can be written
by applying the duality transformations $\overrightarrow{E}\longrightarrow\overrightarrow{H}$,$\overrightarrow{H}\longrightarrow-\overrightarrow{E}$
and $e_{1}\longmapsto g_{1}$ as,

\begin{eqnarray}
\overrightarrow{F_{g}} & = & g_{1}\left[\overrightarrow{H}-\overrightarrow{v}\times\overrightarrow{E}\right].\label{eq:5}\end{eqnarray}
Schwinger-Zwangiger\cite{key-2} generalized equations (\ref{eq:1})
and (\ref{eq:5}) as the equation of motion for the force experienced
by a particle (dyon) carrying simultaneous existence of electric and
magnetic charges as

\begin{align}
\overrightarrow{F} & =M\,\frac{d\overrightarrow{v}}{dt}=\overrightarrow{F_{e}}+\overrightarrow{F_{g}}=e_{1}\left[\overrightarrow{E}+\overrightarrow{v}\times\overrightarrow{H}\right]+g_{1}\left[\overrightarrow{H}-\overrightarrow{v}\times\overrightarrow{E}\right]\label{eq:6}\end{align}
where we may write the following forms of the electric and magnetic
field strengths at the point with $\overrightarrow{r}$ of magnitude
$r$,

\begin{align}
\overrightarrow{E} & =e_{2}\frac{\overrightarrow{r}}{r^{3}},\,\, and\,\,\,\,\,\overrightarrow{H}=g_{2}\frac{\overrightarrow{r}}{r^{3}}\label{eq:7}\end{align}
in which the electric $e_{2}$ and magnetic $g_{2}$ charges for a
stationary body are located at origin. Substituting $\overrightarrow{E}$
and $\overrightarrow{H}$ from equation (\ref{eq:7}) into the equation
(\ref{eq:6}), we get the following expression for equation of motion
for Schwinger-Zwangiger \cite{key-2} dyon as

\begin{align}
\overrightarrow{F} & =M\,\frac{d\overrightarrow{v}}{dt}=\alpha_{12}\frac{\overrightarrow{r}}{r^{3}}+\beta_{12}\frac{\overrightarrow{r}\times(\overrightarrow{v}\times\overrightarrow{r})}{r^{3}}\label{eq:8}\end{align}
where 

\begin{align}
\alpha_{12}= & (e_{1}e_{2}+g_{1}g_{2})\,\,\,\,\, and\,\,\,\,\,\beta_{12}=(e_{1}g_{2}-g_{1}e_{2})\label{eq:9}\end{align}
are respectively known as electric and magnetic coupling parameters
of dyons. Taking the vector product of equation (\ref{eq:8}) with
$\overrightarrow{r}$, we get 

\begin{eqnarray}
\overrightarrow{r}\times M\frac{d\overrightarrow{v}}{dt} & = & \alpha_{12}\frac{\overrightarrow{(r}\times\overrightarrow{r)}}{r^{3}}+\beta_{12}\frac{\overrightarrow{r}\times(\overrightarrow{v}\times\overrightarrow{r})}{r^{3}}\label{eq:10}\end{eqnarray}
and using vector multiplication rule $\overrightarrow{(r}\times\overrightarrow{r)}=0$
and $\frac{d}{dt}(\overrightarrow{r}\times M\overrightarrow{v})=\overrightarrow{r}\times M\frac{d\overrightarrow{v}}{dt}$,
the equation (\ref{eq:10}) reduces to ,

\begin{flushleft}
\begin{equation}
\frac{d}{dt}(\overrightarrow{r}\times M\overrightarrow{v})=\beta_{12}\frac{\overrightarrow{r}\times(\overrightarrow{v}\times\overrightarrow{r})}{r^{3}}=\beta_{12}\frac{d}{dt}(\frac{\overrightarrow{r}}{r})\label{eq:11}\end{equation}
where we have used the identity $\frac{\overrightarrow{r}\times(\overrightarrow{v}\times\overrightarrow{r})}{r^{3}}=\frac{d}{dt}(\frac{\overrightarrow{r}}{r})$.
So that we may define following expression for conserved angular momentum
for dyon as
\par\end{flushleft}

\begin{align}
\overrightarrow{J}= & \overrightarrow{r}\times M\overrightarrow{v}-\beta_{12}\frac{\overrightarrow{r}}{r}\label{eq:12}\end{align}
which gives rise the component of the angular momentum $\overrightarrow{J}$
along the direction of $\overrightarrow{r}$ as

\begin{alignat}{1}
J_{r}=\hat{r}\cdot\overrightarrow{J} & =\beta_{12}\hat{r}\cdot\frac{\overrightarrow{r}}{r}=\beta_{12}\label{eq:13}\end{alignat}
where $\hat{r}$ is the unit vector along the vector $\overrightarrow{r}$
and we have used $\overrightarrow{(r}\times\overrightarrow{r)}=0$.
Thus the quantization of the component of the angular momentum $J_{r}$
along the line of particle leads to Schwinger-Zwangiger\cite{key-2}
charge quantization ( or chirality quantization) in the units of Plank's
constant 

\begin{align}
\beta_{12}= & (e_{1}g_{2}-g_{1}e_{2})=\nu\label{eq:14}\end{align}
where $\nu$ may be an integer or half integer. But in most cases,
it has been taken as integer as its half integral values were already
excluded by Dirac in his seminal paper\cite{key-1}. On substituting
$e_{1}=e;\,\, g_{1}=0;\,\, e_{2}=0$ and $g_{2}=g$ in equation (\ref{eq:14})
for interaction of two dyons with charges $(e,0)$ and $(0,g)$, we
get the Dirac quantization condition\cite{key-1} 

\begin{align}
eg & =\nu\label{eq:15}\end{align}
in the units of Plank constant. It is clear that if we do not consider
dyon Dirac quantization condition is not dual invariant.

\section{Schwinger-Zwanziger Dyon in Gravito-Heavisidian Fields}

	Analogy between electromagnetic and linear gravitational field equations
suggests \cite{key-7} the structural symmetry between these two forces
of nature. Accordingly, on postulating the existence of Heavisidian
monopole \cite{key-11} leads the symmetry between the linear equations
of gravito-Heavisidian fields and the generalized Maxwell's Dirac
equations in electromagnetic Fields. Likewise, generalized Maxwell's
Dirac equations (\ref{eq:2}), we may write the the following form
\cite{key-8,key-11} of linear equations for generalized gravito-
Heavisidian fields in presence of Heavisidian monopole as 

\begin{align}
\overrightarrow{\nabla}\cdot\overrightarrow{\mathsf{G}} & =-\rho_{g}\nonumber \\
\overrightarrow{\nabla}\cdot\overrightarrow{\mathsf{H}}=- & \rho_{h}\nonumber \\
\overrightarrow{\nabla}\times\overrightarrow{\mathsf{G}} & =-\frac{\partial\overrightarrow{\mathsf{H}}}{\partial t}+\overrightarrow{j_{h}}\nonumber \\
\overrightarrow{\nabla}\times\overrightarrow{\mathsf{H}}= & \overrightarrow{j_{g}}+\frac{\partial\overrightarrow{\mathsf{G}}}{\partial t}\label{eq:16}\end{align}
where $\overrightarrow{\mathsf{G}}$ is the gravitational (gravi-electric)
field, $\overrightarrow{\mathsf{H}}$ is the Heavisidian (gravi-magnetic)
field, $\rho_{g}$ is the gravitational charge (mass) density, $\rho_{h}$
is Heavisidian charge (mass) density, $\overrightarrow{j_{g}}$ is
gravitational current density and $\overrightarrow{j_{h}}$ denotes
the Heavisidian current density. Similarly, the force $\overrightarrow{F_{m}}$
the force experienced by a gravitational (gravi-electric) charge (mass)
$m_{1}$ may then be expressed as ,

\begin{eqnarray}
\overrightarrow{F_{m}} & =m_{1}\frac{d\overrightarrow{v}}{dt} & =m_{1}\left[\overrightarrow{\mathsf{G}}-\overrightarrow{v}\times\overrightarrow{\mathcal{\mathsf{H}}}\right].\label{eq:17}\end{eqnarray}
Accordingly GDM type equations of gravito-Heavisidian fields are invariant
not only under Lorentz and conformal transformations but also invariant
under the following duality transformations between gravitational
(gravi-electric) $\mathcal{G}$ and Heavisidian (gravi-magnetic)$\mathcal{M}$
quantities i.e

\begin{align}
\mathcal{G}\Longrightarrow & \mathcal{G}\cos\vartheta+\mathcal{\mathcal{M}\sin\vartheta}\nonumber \\
\mathcal{M}\Longrightarrow & \mathcal{M}\cos\vartheta-\mathcal{\mathcal{G}\sin\vartheta}\label{eq:18}\end{align}
where $\mathcal{G}=\left(m_{1},\overrightarrow{\mathcal{\mathsf{G}}},\rho_{g},\overrightarrow{j_{g}}\right)$
and $\mathcal{\mathcal{M}=}\left(h_{1},\overrightarrow{\mathsf{H}},\rho_{h},\overrightarrow{j_{h}}\right)$
with $h_{1}$denotes the Heavisidian (gravi-magnetic) charge (mass).
For a particular value of $\vartheta=\frac{\pi}{2}$, equations (\ref{eq:18})
reduces to

\begin{align}
\mathcal{G}\longmapsto & \mathcal{M}\,\,\,\,\,\,\mathcal{M}\longmapsto-\mathcal{G}.\label{eq:19}\end{align}
So, on applying the duality symmetry on gravitational and Heavisidian
fields and charges (masses) like the electromagnetism, we may write
the net force experienced by Heavisidian (gravi-magnetic) charge (mass)
$h_{1}$ as ,

\begin{eqnarray}
\overrightarrow{F_{h}} & =h_{1}\frac{d\overrightarrow{v}}{dt}= & h_{1}\left[\overrightarrow{\mathsf{H}}+\overrightarrow{v}\times\overrightarrow{G}\right].\label{eq:20}\end{eqnarray}
Following Schwinger-Zwangiger\cite{key-2}, here also we may adopt
the same process for gravito-dyons. Hence the gravito dyons are considered
as the particles carrying simultaneously the existence of gravitational
(gravi-electric) and Heavisidian (gravi-magnetic) charges (masses).
Thus, we may write the net force $\overrightarrow{\mathsf{F}}$ acting
on gravito-dyon \cite{key-11} as 

\begin{align}
\overrightarrow{\mathsf{F}}= & M\,\frac{d\overrightarrow{v}}{dt}=(m_{1}+h_{1})\frac{d\overrightarrow{v}}{dt}=m_{1}\left[\overrightarrow{\mathsf{G}}-\overrightarrow{v}\times\overrightarrow{\mathcal{\mathsf{H}}}\right]+h_{1}\left[\overrightarrow{\mathsf{H}}+\overrightarrow{v}\times\overrightarrow{G}\right]\label{eq:21}\end{align}
where following expressions for stationary gravitational (gravi-electric)
and Heavisidian (gravi-magnetic) field strengths at the point with
$\overrightarrow{r}$ of magnitude $r$ may be used with gravitational
charge(mass) $m_{2}$ and Heavisidian charge(mass) $h_{2}$ as

\begin{align}
\overrightarrow{\mathsf{G}}= & m_{2}\frac{\overrightarrow{r}}{r^{3}}\,\,\: and\,\,\,\,\,\overrightarrow{\mathsf{H}}=h_{2}\frac{\overrightarrow{r}}{r^{3}}.\label{eq:22}\end{align}
Substituting $\overrightarrow{\mathcal{\mathsf{G}}}$ and $\overrightarrow{\mathsf{H}}$
from equation (\ref{eq:22}) into the equation (\ref{eq:21}), we
get the following expression for equation of motion for Schwinger-Zwangiger
\cite{key-2} gravito-dyon as

\begin{align}
\overrightarrow{\mathsf{F}} & =M\,\frac{d\overrightarrow{v}}{dt}=(m_{1}+h_{1})\frac{d\overrightarrow{v}}{dt}=\gamma_{12}\frac{\overrightarrow{r}}{r^{3}}+\delta_{12}\frac{\overrightarrow{r}\times(\overrightarrow{v}\times\overrightarrow{r})}{r^{3}}\label{eq:23}\end{align}
where \begin{align}
\gamma_{12}= & (m_{1}m_{2}+h_{1}h_{2})\,\,\,\,\, and\,\,\,\,\,\delta_{12}=(m_{1}h_{2}-h_{1}m_{2})\label{eq:24}\end{align}
are respectively known as and gravitational (gravi-electric) and Heavisidian
(gravi-magnetic) coupling parameters of gravito-dyons. Taking the
vector product of equation (\ref{eq:23}) with $\overrightarrow{r}$,
we get 

\begin{eqnarray}
\overrightarrow{r}\times(m_{1}+h_{1})\frac{d\overrightarrow{v}}{dt} & = & \gamma_{12}\frac{\overrightarrow{(r}\times\overrightarrow{r)}}{r^{3}}+\delta_{12}\frac{\overrightarrow{r}\times(\overrightarrow{v}\times\overrightarrow{r})}{r^{3}}.\label{eq:25}\end{eqnarray}
Using vector multiplication rule $\overrightarrow{(r}\times\overrightarrow{r)}=0$
and $\frac{d}{dt}\left[\overrightarrow{r}\times(m_{1}+h_{1})\overrightarrow{v}\right]=\overrightarrow{r}\times(m_{1}+h_{1})\frac{d\overrightarrow{v}}{dt}$,
the equation (\ref{eq:25}) reduces to ,

\begin{flushleft}
\begin{equation}
\frac{d}{dt}\left[\overrightarrow{r}\times(m_{1}+h_{1})\overrightarrow{v}\right]=\delta_{12}\frac{\overrightarrow{r}\times(\overrightarrow{v}\times\overrightarrow{r})}{r^{3}}=\delta_{12}\frac{d}{dt}(\frac{\overrightarrow{r}}{r})\label{eq:26}\end{equation}
where $\frac{\overrightarrow{r}\times(\overrightarrow{v}\times\overrightarrow{r})}{r^{3}}=\frac{d}{dt}(\frac{\overrightarrow{r}}{r})$.
As such, we may write the following expression for conserved angular
momentum for gravito-dyon as
\par\end{flushleft}

\begin{align}
\overrightarrow{\mathcal{J}}= & \overrightarrow{r}\times(m_{1}+h_{1})\overrightarrow{v}-\delta_{12}\frac{\overrightarrow{r}}{r}\label{eq:27}\end{align}
which gives rise the component of the angular momentum $\mathcal{\overrightarrow{J}}$
along the direction of $\overrightarrow{r}$ as

\begin{alignat}{1}
\mathcal{J}_{r}=\hat{r}\cdot\overrightarrow{J} & =\delta_{12}\hat{r}\cdot\frac{\overrightarrow{r}}{r}=\delta_{12}\label{eq:28}\end{alignat}
where $\hat{r}$ is the unit vector along the vector $\overrightarrow{r}$
and $\overrightarrow{(r}\times\overrightarrow{r)}=0$. Thus the quantization
of the component of the angular momentum $J_{r}$ along the line of
particle leads to Schwinger-Zwangiger \cite{key-2} charge (mass)
quantization ( or chirality quantization) condition for gravito-dyons
in the units of Plank's constant 

\begin{align}
\beta_{12}= & (e_{1}g_{2}-g_{1}e_{2})=\mathtt{n}\label{eq:29}\end{align}
where $\mathtt{n}$ may be an integer or half integer as may be the
case of dyons in electromagnetic fields.. On substituting $m_{1}=m;\,\, h_{1}=0;\,\, m_{2}=0$
and $h_{2}=h$ in equation (\ref{eq:29}) for interaction of two gravito-dyons
with charges (masses) $(m,0)$ and $(0,h)$, we get the Dirac quantization
condition\cite{key-1} for gravito-dyons as 

\begin{align}
mh & =\mathtt{n}\label{eq:30}\end{align}
which can be dual invariant only if we consider the case of gravito-dyons
in linear gravitational fields.

\section{Generalization Schwinger-Zwanziger Dyon to Quaternion}

Let us assume that a particle of mass $M$ carries simultaneous existence
of four charges namely electric ($e_{1}$), magnetic ($g_{1}$), gravitational
(gravi-electric) $(m_{1})$ and Heavisidian (gravi-magnetic) $(h_{1})$.
Let this particle moves with velocity $\overrightarrow{v}$ so that
it experiences a combined force which is the sum of the forces exerted
independently due to individual charges i.e. 

\begin{align}
\boldsymbol{\overrightarrow{F}}= & \overrightarrow{F_{e}}+\overrightarrow{F_{g}}+\overrightarrow{F_{m}}+\overrightarrow{F_{h}}\label{eq:31}\end{align}
where $\overrightarrow{F_{e}}$ , $\overrightarrow{F_{g}}$, $\overrightarrow{F_{g}}$
and $\overrightarrow{F_{h}}$ are respectively given by equations
(\ref{eq:1}), (\ref{eq:5}), (\ref{eq:17}) and (\ref{eq:20}). Now
substituting the values of electric $(\overrightarrow{E})$, magnetic
($\overrightarrow{H}$), gravitational (gravi-electric) $(\mathsf{\overrightarrow{G}})$
and Heavisidian (gravi-magnetic) $(\mathsf{\overrightarrow{H}})$
field strengths given by equations (\ref{eq:7}), and (\ref{eq:8})
in to equation (\ref{eq:31}), we get 

\begin{align}
\boldsymbol{\overrightarrow{F}}= & W_{12}\frac{\overrightarrow{r}}{r^{3}}+(X_{12}+Y_{12}+Z_{12})\overrightarrow{v}\times\frac{\overrightarrow{r}}{r^{3}}\label{eq:32}\end{align}
where 

\begin{align}
W_{12}= & (e_{1}e_{2}+g_{1}g_{2}+m_{1}m_{2}+h_{1}h_{2});\label{eq:33}\\
X_{12}= & (e_{1}g_{2}-e_{2}g_{1}+m_{1}h_{2}-m_{2}h_{1});\label{eq:34}\\
Y_{12}= & (e_{1}m_{2}-m_{1}e_{2}-h_{2}g_{1}+h_{1}g_{2});\label{eq:35}\\
Z_{12}= & (e_{1}h_{2}-h_{1}e_{2}+g_{1}m_{2}-m_{1}g_{2}).\label{eq:36}\end{align}
are different four coupling parameters. $W_{12},$$X_{12},$$Y_{12}$,$Z_{12}$
may also be identified as electric, magnetic, gravitational and Heavisidian
parameters \cite{key-9}. According to the Newton's second law $\overrightarrow{F}=M\frac{d\overrightarrow{v}}{dt}$
, equation (\ref{eq:32}) is written as 

\begin{eqnarray}
M\frac{d\overrightarrow{v}}{dt} & = & W_{12}\frac{\overrightarrow{r}}{r^{3}}+(X_{12}+Y_{12}+Z_{12})\overrightarrow{v}\times\frac{\overrightarrow{r}}{r^{3}}.\label{eq:37}\end{eqnarray}
Taking the vector product of equation (\ref{eq:37}) with $\overrightarrow{r}$,
we get 

\begin{eqnarray}
\overrightarrow{r}\times M\frac{d\overrightarrow{v}}{dt} & = & W_{12}\frac{\overrightarrow{(r}\times\overrightarrow{r)}}{r^{3}}+(X_{12}+Y_{12}+Z_{12})\frac{\overrightarrow{r}\times(\overrightarrow{v}\times\overrightarrow{r})}{r^{3}}.\label{eq:38}\end{eqnarray}
Using vector multiplication rule $\overrightarrow{(r}\times\overrightarrow{r)}=0$
, $\frac{d(\overrightarrow{r}\times M\overrightarrow{v)}}{dt}=\overrightarrow{r}\times m\frac{d\overrightarrow{v}}{dt}$,
and the identity $\frac{\overrightarrow{r}\times(\overrightarrow{v}\times\overrightarrow{r})}{r^{3}}=\frac{d}{dt}(\frac{\overrightarrow{r}}{r})$
, equation (\ref{eq:38}) reduces to ,

\begin{flushleft}
\begin{equation}
\frac{d}{dt}(\overrightarrow{r}\times M\overrightarrow{v})=(X_{12}+Y_{12}+Z_{12})\frac{d}{dt}(\frac{\overrightarrow{r}}{r}).\label{eq:39}\end{equation}
Thus by adopting the above procedure, we get the following expression
for angular momentum $\mathsf{\overrightarrow{J}}$ as 
\par\end{flushleft}

\begin{flushleft}
\begin{equation}
\mathsf{\overrightarrow{J}}=\overrightarrow{r}\times M\,\overrightarrow{v}-(X_{12}+Y_{12}+Z_{12})(\frac{\overrightarrow{r}}{r})\label{eq:40}\end{equation}
which gives rise the component of the angular momentum $\overrightarrow{\mathsf{J}}$
along the direction of $\overrightarrow{r}$ as,
\par\end{flushleft}

\begin{eqnarray}
\mathsf{J}_{r} & = & (X_{12}+Y_{12}+Z_{12}).\label{eq:41}\end{eqnarray}
This is called the residual component of unified angular momentum
and leads the following form of generalized Schwinger-Zwangiger \cite{key-2}
quantization condition i.e \begin{eqnarray}
\mathsf{J}_{r}= & (X_{12}+Y_{12}+Z_{12}) & =n\label{eq:42}\end{eqnarray}
where $n$ is an integer and $h$ is Planck's constant.

\section{Quaternion Formulation for Gravi-Electromagnetism}

		A unified theory of generalized electromagnetic and Heavisidian
fields may then be developed consistently by generalizing Schwinger-Zwangiger
\cite{key-2} dyon to a quaternion possessing a quartet $(e,\, g,\, m,\, h)$
of four charges. Quaternion $(e,\, g,\, m,\, h)$ charge thus represents
the theory of gravi-electromagnetism for the particles carrying simultaneously
electric, magnetic, gravitational and Heavisidian charges. So, let
us generalize two types of Schwinger-Zwangiger \cite{key-2} dyonic
charges $(e,\, g)$ and $(m,\, h)$ to a quaternion charge of gravi-
electromagnetic fields as \cite{key-8} as,

\begin{align}
Q=(e-ig)-j(m-ih)= & e-ig-jm-kh\label{eq:43}\end{align}
where $i,\, j,\, k$ are three quaternion basis elements satisfying
the quaternion multiplication rules\cite{key-8}

\begin{align}
i\cdot i=j\cdot j=k\cdot k & =-1\nonumber \\
i\cdot j=-j\cdot i & =k\nonumber \\
j\cdot k=-k\cdot j & =i\nonumber \\
k\cdot i=-i\cdot k & =j.\label{eq:44}\end{align}
The quaternion conjugation of equation (\ref{eq:43}) is defined as

\begin{align}
\overline{Q}=(e+ig)+(m+ih)j= & e+ig+jm+kh.\label{eq:45}\end{align}
Thus, on using the quaternion multiplication rule (\ref{eq:44}),
the interaction of between two quaternions ($a$ and $b$ ) with charges
$Q_{a}=(e_{a},g_{a},m_{a},h_{a})=e_{a}-ig_{a}-jm_{a}-kh_{a}$ and
$Q_{b}=(e_{b},g_{b},m_{b},h_{b})=e_{b}-ig_{b}-jm_{b}-kh_{b}$ leads
\cite{key-9} to

\begin{align}
\overline{Q_{a}}Q_{b} & =\left(e_{a}+ig_{a}+jm_{a}+kh_{a}\right)\left(e_{b}-ig_{b}-jm_{b}-kh_{b}\right)\nonumber \\
= & \alpha_{ab}+\beta_{ab}+\gamma_{ab}+\delta_{ab}\label{eq:46}\end{align}
where

\begin{align}
\alpha_{ab}= & (e_{a}e_{b}+g_{a}g_{b}+m_{a}m_{b}+h_{a}h_{b});\label{eq:47}\\
\beta_{ab}= & (e_{a}g_{b}-e_{b}g_{a}+m_{a}h_{b}-m_{b}h_{a});\label{eq:48}\\
\gamma_{ab}= & (e_{a}m_{b}-m_{a}e_{b}-h_{b}g_{a}+h_{a}g_{b});\label{eq:49}\\
\delta_{ab}= & (e_{a}h_{b}-h_{a}e_{b}+g_{a}m_{b}-m_{a}g_{b}).\label{eq:50}\end{align}
Hence,for $a=1$ and $b=2$, equations (\ref{eq:47}- \ref{eq:50})
are same as equations (\ref{eq:33}- \ref{eq:36}) for four different
chirality parameters i.e $\alpha_{12}=W_{12}$,~$\beta_{12}=$$X_{12},$~$\gamma_{12}=$$Y_{12}$,~
and $\delta_{12}=Z_{12}$. So, the equations (\ref{eq:33}) and (\ref{eq:34})
immediately reduces to $W_{12}=e_{1}e_{2}+g_{1}g_{2}$ and $X_{12}=(e_{1}g_{2}-e_{2}g_{1})$
for the interaction of two Schwinger-Zwangiger \cite{key-2} dyons
from two quaternion charges $(e_{1},g_{1},0,0)$ and $(e_{2},g_{2},0,0)$
with the vanishing of other parameters. Similarly, for the Schwinger-Zwangiger
case for interaction of gravito-dyons i.e. gravitational charge and
Heavisidian monopole , we get $W_{12}=m_{1}m_{2}+h_{1}h_{2}$ and
$X_{12}=m_{1}h_{2}-h_{1}m_{2}$ from quaternions $(0,0,m_{1},h_{1})$
and $(0,0,m_{2},h_{2})$. Similarly, we may be speculate a new kind
of dyon i.e. electric charge and Heavisidian monopole obtained from
a quaternion like $\left(e,0,0,h\right)$ where we may get $W_{12}=e_{1}e_{2}+h_{1}h_{2}$
and $Z_{12}=e_{1}h_{2}-h_{1}e_{2}$. Similarly a new kind of dyon
i.e., electric charge and gravitational charge may also be speculated
from the quaternion like $\left(e,0,m,0\right)$ for which we get
$W_{12}=e_{1}e_{2}+m_{1}m_{2}$ and $Y_{12}=e_{1}m_{2}-m_{1}e_{2}.$
Also there are the possibilities of other dyons like magnetic charge
and gravitational charge from a quaternion $\left(0,g,m,0\right)$
for which $W_{12}=m_{1}m_{2}+g_{1}g_{2}$ and $Z_{12}=g_{1}m_{2}-m_{1}g_{2}$
as well as for the purely hypothetical dyons like magnetic and Heavisidian
monopoles we get $W_{12}=h_{1}h_{2}+g_{1}g_{2}$\textbf{ }and $Y_{12}=h_{1}g_{2}-g_{1}h_{2}$
. As such, the quaternion generalization of Schwinger-Zwangiger \cite{key-2}
quantization condition of dyons extends the possibilities of six types
of dyons like $(e,\, g)$, $(m,\, h)$, $(e,\, h)$, $(e,h)$, $(g,\, m)$
and $(g,\, h)$. Let us try to obtain the six kinds of Schwinger-Zwangiger
dyons from a quaternion (\ref{eq:43}).
\begin{itemize}
\item Applying the $j-$conjugation i.e. $i\rightarrow i$,$j\rightarrow-j$
, $k\rightarrow-k$ as $k=ij$ on quaternion (\ref{eq:43}), we get,
\end{itemize}
\begin{flushleft}
\begin{eqnarray}
q^{j} & = & e-ig+jm+kh\label{eq:51}\end{eqnarray}
and adding this equations (\ref{eq:51}) to quaternion (\ref{eq:43}),
we get 
\par\end{flushleft}

\begin{flushleft}
\begin{eqnarray}
e-ig & = & \frac{1}{2}(q+q^{j})\label{eq:52}\end{eqnarray}
which refers to generalized charge (as complex quantity) as the order
pair of $(e,\, g)$ for Schwinger-Zwangiger dyon moving in electromagnetic
fields.
\par\end{flushleft}
\begin{itemize}
\item \begin{flushleft}
Similarly, on subtracting equation (\ref{eq:51}) from equation (\ref{eq:43}),
we get 
\par\end{flushleft}
\end{itemize}
\begin{flushleft}
\begin{equation}
m-ih=\frac{1}{2j}(q-q^{j})\label{eq:53}\end{equation}
which refers to generalized charge (as complex quantity) as the order
pair of $(m,\, h)$ for Schwinger-Zwangiger dyon moving in gravito-Heavisidian
fields.
\par\end{flushleft}
\begin{itemize}
\item \begin{flushleft}
Now, applying $i-$conjugation i.e $i\rightarrow-i$ ,$j\rightarrow j$
, $k\rightarrow-k$ as $k=ij$ on quaternion (\ref{eq:43}), we get 
\par\end{flushleft}
\end{itemize}
\begin{flushleft}
\begin{eqnarray}
q^{i} & = & e+ig-jm+kh\label{eq:54}\end{eqnarray}
and adding the this equations (\ref{eq:54}) to (\ref{eq:43}) we
find 
\par\end{flushleft}

\begin{flushleft}
\begin{eqnarray}
e-jm & = & \frac{1}{2}(q+q^{j})\label{eq:55}\end{eqnarray}
which defines the generalized charge (as complex quantity) as the
order pair of $(e,\, m)$ for of a Schwinger-Zwangiger dyon obtained
from electric and gravitational charges.
\par\end{flushleft}
\begin{itemize}
\item \begin{flushleft}
Furthermore, on subtracting equation (\ref{eq:25}) from equation
(\ref{eq:21}), we get 
\par\end{flushleft}
\end{itemize}
\begin{flushleft}
\begin{eqnarray}
(g-jh) & = & \frac{1}{2\, i}(q-q^{i})\label{eq:56}\end{eqnarray}
which describes the generalized charge (as complex quantity) as the
order pair of $(g,\, h)$ for of a Schwinger-Zwangiger dyon obtained
from magnetic and Heavisidian charges.
\par\end{flushleft}
\begin{itemize}
\item \begin{flushleft}
Applying the transformation for $i\rightarrow-i$,\ $j\rightarrow j$,
$k\rightarrow-k$ in equation (\ref{eq:43}), we get,
\par\end{flushleft}
\end{itemize}
\begin{flushleft}
\begin{eqnarray}
q^{k} & = & e+ig+jm-kh\label{eq:57}\end{eqnarray}

\par\end{flushleft}

\begin{flushleft}
and adding equations (\ref{eq:43}) and (\ref{eq:57}), we get 
\par\end{flushleft}

\begin{flushleft}
\begin{eqnarray}
e-kh & = & \frac{1}{2}(q+q^{k}).\label{eq:58}\end{eqnarray}

\par\end{flushleft}

\begin{flushleft}
which gives the generalized charge (as complex quantity) as the order
pair of $(e,\, h)$ for of a Schwinger-Zwangiger dyon obtained from
from electric and Heavisidian charges.
\par\end{flushleft}
\begin{itemize}
\item \begin{flushleft}
Similarly on subtracting the equation (\ref{eq:57}) from equation
(\ref{eq:43}) , we find,
\par\end{flushleft}
\end{itemize}
\begin{flushleft}
\begin{eqnarray}
q-q^{k} & = & -2(ig+jm)\label{eq:59}\end{eqnarray}
and applying the quaternion property $j=-ik$, the equation (\ref{eq:59})
reduces to ,
\par\end{flushleft}

\begin{flushleft}
\begin{eqnarray}
(g-km) & = & \frac{i(q-q^{k})}{2}\label{eq:60}\end{eqnarray}
which describes the generalized charge (as complex quantity) as the
order pair of $(g,\, m)$ for of a Schwinger-Zwangiger dyon obtained
from magnetic and gravitational charges.
\par\end{flushleft}

\textbf{Acknowledgment: }One of us OPSN expresses his thanks of gratitude
to German Academic Exchange Service (Deutscher Akademischer Austausch
Dienst), Bonn for their financial support under DAAD re-invitation
programme at Universität Konstanz.

\end{document}